\documentstyle[12pt]{article}
\begin{document}
\begin{titlepage}

\centerline{\large \bf 
Nonmonotonic external field dependence of the magnetization } 
\centerline{\large \bf 
in a finite Ising model: theory and MC simulation} 
\vspace*{0.4cm}
\centerline{X.S. Chen$^{1,2,a}$, V. Dohm$^{2,b}$ and D. Stauffer$^{3,c}$}
\centerline{$^1$ Institute of Particle Physics, Hua-Zhong 
Normal University,}
\centerline{Wuhan 430079, P.R. China}
\centerline{$^2$ Institut f\"{u}r Theoretische Physik, Technische Hochschule 
Aachen,}
\centerline{D-52056 Aachen, Germany}
\centerline{$^3$ Institute for Theoretical Physics, Cologne University,}
\centerline{D-50923 K\"oln, Germany}
\vspace*{0.3cm}

\begin{abstract}

Using $\varphi^4$ field theory and Monte Carlo (MC) simulation we investigate
the finite-size effects of the magnetization $M$ for the three-dimensional 
Ising model in a finite cubic geometry with periodic boundary conditions. 
The field theory with infinite cutoff gives a scaling form of 
the equation of state $h/M^\delta = f(hL^{\beta\delta/\nu},
t/h^{1/\beta\delta})$ where $t=(T-T_c)/T_c$ is the reduced temperature, 
$h$ is the external field and $L$ is the size of system. Below $T_c$ and 
at $T_c$ the theory predicts a nonmonotonic dependence of $f(x,y)$ with 
respect to $x \equiv hL^{\beta\delta/\nu}$ at fixed $y \equiv t/h^{1/\beta
\delta}$ and a crossover from nonmonotonic to monotonic behaviour 
when $y$ is further increased. These results are confirmed by MC simulation. 
The scaling function $f(x,y)$ obtained from the field theory is in good 
quantitative agreement with the finite-size MC data. Good agreement is also
found for the bulk value $f(\infty,0)$ at $T_c$.

\end{abstract}

PACS: 05.70.Jk, 64.60.-i

$^a$ e-mail: chen@physik.rwth-aachen.de  
\newline
$^b$ e-mail: vdohm@physik.rwth-aachen.de
\newline
$^c$ e-mail: stauffer@thp.uni-koeln.de
\end{titlepage}

\section{ Introduction}

The scaling equation of state near a critical point provides fundamental
information on the critical behaviour of a thermodynamic system. For bulk
Ising-like systems accurate predictions have been made recently \cite{zinn}
on the basis of the $\varphi^4$ field theory in three dimensions. Testing 
these predictions by Monte Carlo simulations \cite{binder} would be of 
considerable interest. Such
simulations, however, are necessarily made only for finite systems and thus 
phenomenological concepts like finite-size scaling \cite{fisher} are 
needed to perform extrapolations from mesoscopic lattices.

In order to test the theory in a more conclusive way it is desirable to go
beyond the phenomenological finite-size scaling concept and to calculate 
explicitly the finite-size effects on the equation of state, i.e., on the 
magnetization as a function of the temperature $T$ and external field $h$
in a finite geometry. Although such field-theoretic calculations 
\cite{BZ,RGJ,EDC,CD1} are 
perturbative and not exact they have been found to be in good agreement with
the MC simulations. So far the calculations of thermodynamic quantities were 
restricted to the case of zero external field $h$. At finite $h$, only the
order-parameter distribution function was calculated \cite{CD0,CDi,rudnick}.
In the present paper we extend these calculations to the equation of state 
of the three-dimensional Ising model in a finite cubic geometry at finite 
$h$. For simplicity these
calculations are performed at infinite cutoff and thus our results neglect
lattice effects. The latter have been shown \cite{CD,CD2} to yield 
(exponentially small) non-universal non-scaling contributions. Here we focus 
our interest on the universal scaling part of the equation of state. Our theory
predicts non-monotonic effects in the $h$ dependence of the equation of state 
which we then confirm with surprisingly good agreement by standard Monte 
Carlo simulations.

The identification of such non-monotonic effects is of great practical 
importance. If, for example,
the critical temperature $T_c(L)$ of a lattice with $L^3$ sites in three 
dimensions varies asymptotically as $T_c(L)-T_c(\infty) \propto 1/L^y$ with
some correlation length exponent $y = 1/\nu$, then a plot of the numerical
$T_c(L)$ versus $1/L^y$ gives the extrapolated $T_c(\infty)$ as an intercept.
If, however, higher order terms make the curve $T_c(L)$ non-monotonic, then
such a plot for finite $L$, where the non-monotonicity is not yet visible,
would give a wrong estimate for $T_c(\infty)$. Similar effects would make
estimates of other quantities unreliable, and there exist examples 
of such estimates in the literature.

\section{Field-theoretic calculations}

We consider the $\varphi^4$ model with the standard Landau-Ginzburg-Wilson 
Hamiltonian
\begin{equation}
H(h)=\int_V \left [ {1\over 2}r_0 \varphi^2 + {1\over 2}
(\bigtriangledown \varphi)^2 + u_0 \varphi^4 - h\cdot \varphi \right ]
\end{equation}
where $\varphi ({\bf x})$ is a one-component field in a finite cube of volume 
$V = L^d$ and $h$ is a homogeneous external field. We assume periodic boundary
conditions. Accordingly we have
\begin{equation}
\varphi ({\bf x}) = L^{-d} \sum_{\bf k} \varphi_{\bf k} e^{i 
{\bf k}\cdot {\bf x} }
\end{equation}
where the summation $\sum_{\bf k}$ runs over discrete 
${\bf k}= {2\pi\over L}{\bf m}$ vectors with componets 
$k_j = {2\pi\over L} m_j, m_j = 0,\pm 1,\pm 2, ..., j= 1,2,...,d$ in the 
range $ -\Lambda \leq k_j < \Lambda$, i.e., with a sharp cutoff $\Lambda$. 
The temperature enters through $r_0 = r_{0c}+ a_0 t, t=(T-T_c)/T_c$. 

As pointed out recently 
\cite{CD2} the Hamiltonian (1) for periodic boundary conditions with a sharp
cutoff $\Lambda$ does not correctly describe the exponential size dependence
of physical quantities of finite lattice models in the region $\xi \gg L$.
Instead of (1), a modified continuum Hamiltonian with a smooth cutoff 
\cite{CD2} would be more appropriate. Even better would be to employ
the lattice version of the $\varphi^4$ theory to describe the non-scaling
lattice effects of finite Ising models in the region $\xi \gg L$. In the 
present paper, however, we shall neglect such effects by taking the limit
$\Lambda \rightarrow \infty$ (see below).

The fluctuating homogeneous part of the order-parameter of the Hamiltonian (1) 
is $\Phi = V^{-1} \int_V d^d x \varphi ({\bf x}) = L^{-d} \varphi_0$. 
As previously \cite{BZ,RGJ} $\varphi$ is decomposed as
\begin{equation}
\varphi ({\bf x}) = \Phi + \sigma ({\bf x})
\end{equation}
where $\sigma ({\bf x})$ includes all inhomogeneous modes
\begin{equation}
\sigma ({\bf x}) = L^{-d} \sum_{{\bf k}\neq {\bf 0}} \varphi_{\bf k}
e^{i {\bf k}\cdot {\bf x}}.
\end{equation}
The order-parameter distribution function $P (\Phi) \equiv P (\Phi,t,h,L)$
is defined by functional integration over $\sigma$,
\begin{equation}
P (\Phi,t,h,L) = Z(h)^{-1}\int D \sigma e^{-H(h)},
\end{equation}
where
\begin{equation}
Z(h) = \int_{-\infty}^{\infty}  d \Phi \int D \sigma e^{-H(h)}
\end{equation}
is the partition function of system. This distribution function depends also 
on the cutoff $\Lambda$ which implies non-scaling finite-size effects 
\cite{CD,CD2}. From the order-parameter distribution 
function $P (\Phi)$ we can calculate the magnetization  
\begin{equation}
M = <|\Phi|> = \int_{-\infty}^{\infty}  d \Phi |\Phi| P (\Phi).
\end{equation}

The functional integration over $\sigma$ in Eqs. (5) and (6) can only be 
done perturbatively \cite{BZ,RGJ,EDC}. Recently a novel perturbation approach 
was presented \cite{CD1,CD0}. Using this approach the order-parameter 
distribution can be written in the form
\begin{equation}
P(\Phi) = e^{-H^{eff}(\Phi)}/\int_{-\infty}^{\infty} d \Phi e^{-H^{eff}(\Phi)}
\end{equation}
where the (bare) effective Hamiltonian of the Ising-like system reads  
\begin{eqnarray}
H^{eff}(\Phi) &=& H_0 (\Phi,h)-{1\over 2}\sum_{{\bf m} \neq {\bf 0}}
\ln \{ \frac{Z_1 [y_{0{\bf m}}(r_{0L})]}{Z_1 [y_{0{\bf m}}(r_{0})]} \}, \\
H_0 (\Phi,h) &=& L^d ({1\over 2}r_0 \Phi^2+u_0 \Phi^4 - h\cdot \Phi),
\end{eqnarray}
with $r_{0L} = r_0 +12 u_0 \Phi^2$ and $y_{0{\bf m}}(r)= 
(2L^d/3u_0)^{1/2}(r+{4\pi^2\over L^2}{\bf m}^2)$. The function $Z_1[y]$
in Eq.(9) is defined as
\begin{equation}
Z_1 [y] = \int_{0}^{\infty} ds \; s\exp(-{1\over 2}y s^2 -s^4).
\end{equation}

After minimal renormalization at fixed dimension $d < 4$ \cite{SD} in
the limit $\Lambda \rightarrow \infty$, we obtain the finite-size scaling 
form of the order-parameter distribution function \cite{CD1,CD0} 
\begin{equation}
P(\Phi,t,h,L) = L^{\beta/\nu} p(hL^{\beta\delta/\nu},tL^{1/\nu},
\Phi L^{\beta/\nu}),
\end{equation}
where
\begin{equation}
p(x,q,z) = {\exp [-F(x,q,z)]\over \int_{-\infty}^{\infty} d z 
\exp [-F(x,q,z)]},
\end{equation}
with $x=hL^{\beta\delta/\nu}$, $q=tL^{1/\nu}$, 
$z=\Phi L^{\beta/\nu}$ and
\begin{eqnarray}
F(x,q,z) &=& c_2 (x,\hat{q})\hat{z}^2+c_4 (x,\hat{q})\hat{z}^4-x\cdot z 
\nonumber \\
&&-{1\over 2}\sum_{{\bf m} \neq {\bf 0}}
\ln \{ \frac{Z_1 [y_{{\bf m}}(\tilde{r}_{L}(x,\hat{q},\hat{z}))]}
{Z_1 [y_{{\bf m}}(\tilde{r}_{L}(x,\hat{q},0))]} \}.
\end{eqnarray}

Here $\hat{q}=Q^* t(L/\xi_0)^{1/\nu}$ and $\hat{z}=(2Q^*)^\beta (\Phi/A_M)
(L/\xi_0)^{\beta/\nu}$ are dimensionless variables normalized by the 
asymptotic amplitudes $\xi_0$ and $A_M$ of the bulk correlation length
$\xi =\xi_0 t^{-\nu}$ at $h=0$ above $T_c$ and of the bulk order-parameter 
$M_{bulk} = A_M |t|^\beta $ at $h=0$ below $T_c$. The bulk parameter $Q^*$ 
is known  \cite{SD}. The coefficients $c_2 (x,\hat{q})$ and 
$c_4 (x,\hat{q})$ read for $d = 3$
\begin{eqnarray}
c_2 (x,\hat{q}) &=& (64\pi u^*)^{-1}\hat{q}\tilde{\ell}(x,
\hat{q})^{3-(2\beta+1)/\nu}(1+12u^*),\\
c_4 (x,\hat{q}) &=& (256\pi u^*)^{-1}\tilde{\ell}(x,\hat{q})^{3-4\beta/\nu}
(1+36u^*),
\end{eqnarray}
where $u^*$ is the fixed point value of the renormalized coupling \cite{SD}.
In three dimensions we have
\begin{eqnarray}
y_{\bf m}(\tilde{r}_L (x,\hat{q},\hat{z})) &=& [6\pi u^*\tilde{\ell}
(x,\hat{q})]^{-1/2}[\tilde{r}_L (x,\hat{q},\hat{z})\tilde{\ell}
(x,\hat{q})^2+4\pi^2 {\bf m}^2],\\
\tilde{r}_L (x,\hat{q},\hat{z}) &=& \hat{q}\tilde{\ell}(x,\hat{q})^{-1/\nu}
+(3/2)\tilde{\ell}(x,\hat{q})^{-2\beta\nu}\hat{z}^2.
\end{eqnarray}
The auxiliary scaling function $\tilde{\ell}(x,\hat{q})$ of the flow 
parameter is determined by
\begin{eqnarray}
\tilde{\ell}(x,\hat{q})^{3/2} &=& (4\pi u^*)^{1/2}[\tilde{y}(x,\hat{q})+
12\vartheta_2 (\tilde{y}(x,\hat{q}),\hat{x})],\\
\tilde{y}(x,\hat{q}) &=& (4\pi u^*)^{-1/2} \tilde{\ell}
(x,\hat{q})^{3/2-1/\nu} \hat{q},\\
\hat{x} &=& A_M (2Q^*)^{-\beta}\xi_0^{\beta/\nu}(4\pi u^*)^{1/4}\sqrt{8} 
\tilde{\ell}(x,\hat{q})^{\beta/\nu-3/4} x ,
\end{eqnarray}
where
\begin{equation}
\vartheta_2 (\tilde{y},\hat{x}) ={\int_0^{\infty} ds \; s^2 \cosh (\hat{x}s) 
\exp (-{1\over 2}\tilde{y} s^2 - s^4)\over 
\int_0^{\infty} ds  \cosh (\hat{x}s) \exp (-{1\over 2} \tilde{y} s^2 - s^4) }.
\end{equation}

From Eqs. (7) and (12) we obtain the scaling form
\begin{eqnarray}
M(h,t,L) &=& L^{-\beta/\nu} f_M (hL^{\beta\delta/\nu},tL^{1/\nu}), \\
f_M (x,q) &=& {\int_{-\infty}^{\infty} d z |z| \exp [-F(x,q,z)]\over 
\int_{-\infty}^{\infty} d z \exp [-F(x,q,z)]}.
\end{eqnarray}
Correspondingly the asymptotic equation of state for a finite Ising-like 
system in the limit of zero lattice spacing \cite{CD2} can be
written as
\begin{equation}
h/M^\delta = f(hL^{\beta\delta/\nu},t/h^{1/\beta\delta})
\end{equation}
where
\begin{equation}
f(x,y)=x/f_M (x,yx^{1/\beta\delta})^{\delta}.
\end{equation}

In the comparison with the MC data of the simple-cubic (sc) Ising model in 
Section 4, the quantities $h$, $M$ and $L$ are used in a dimensionless 
form in units of the lattice constant, see Sect. 4 of Ref.\cite{CDi}. 
We have taken the bulk parameters $u^*=0.0412, Q^* = 0.945$ from 
Ref.\cite{EDC}. From Ref.\cite{LF} we have taken the bulk amplitudes 
$\xi_0 = 0.495, A_M = 1.71$ in units of the lattice constant 
(of the sc Ising model) and the bulk critical exponents $\beta = 0.3305, 
\nu= 0.6335$. Thus our determination of the scaling function $f(x,y)$ does 
not require a new adjustment of nonuniversal parameters.

Taking the limit $ hL^{\beta\delta/\nu} \rightarrow \infty$ at fixed 
$t/h^{1/\beta\delta}$, we obtain the 
scaling form of the bulk equation of state 
\begin{equation}
h/M^\delta = f_b (t/h^{1/\beta\delta}) = f(\infty,t/h^{1/\beta\delta}).
\end{equation}
At $T = T_c $ we find from Eqs. (23)-(27)
\begin{equation}
h/M^\delta \equiv D_c = f(\infty,0) = 0.202
\end{equation}
in three dimensions (in units of the lattice constant).

\section{Monte Carlo simulation}

Standard heat bath techniques were used for the Glauber kinetic Ising model,
with multi-spin coding (16 spins in each 64-bit computer word). Since the 
effect could be seen best in lattices of intermediate size $L \simeq 80$
for $L \times L \times L$  spins,
memory requirements were tiny and only trivial parallelization by replication,
not by domain decomposition, was used. However, thousands of hours of processor
time were needed since we see the non-monotonic effects clearly in our figure
if $h/M^\delta$ is plotted. The exponent $\delta$ is nearly five and thus
five percent accuracy in $h/M^\delta$ requires one percent accuracy in the 
directly simulated magnetization $M$. 

Earlier simulations by the same author
and algorithm \cite{stauffer} did not show the non-monotonic effects 
at $T=T_c$ because they were not searched for; at that time the 
field-theoretical 
predictions presented above were not yet known. But even if they had been known
it is doubtful that with the Intel Paragon used in \cite{stauffer} instead of
the Cray-T3E now the non-monotonic trends would have been seen in about the
same computer time. 

Because of the limited system size, errors in the magnetization of order
1 + const/$L^{\beta/\nu} \simeq$ 1 + const/$\sqrt{L}$ are expected; in light
of these errors the agreement to be presented now is surprisingly good.

\section{Results and discussion}

Figure 1 compares our Monte Carlo results for $32^3$ and $80^3$ spins with our 
theoretical predictions, Eqs. (23)-(26), and shows good quantitative 
agreement; below $T_c$
a pronounced peak is seen in the scaling function (Fig.1a), at $T_c$ it is
somewhat weaker (Fig.1b), above $T_c$ it is barely visible (Fig.1c), and
far above $T_c$ (Fig.1d) it has vanished. The simulations, which at $T=T_c$
have an accuracy of the order of one and five percent for $L=32$ and 80, 
respectively, agree nicely with the theoretical predictions. (Far above $T_c$ 
therefore no simulations were made.)

Thus, if one tries to determine the bulk critical amplitude of $h/M^\delta$ at
the critical isotherm, then with varying $L$ at a fixed field one first gets
a too small value (left border of the figures), then a too high value (peak),
and then a roughly correct value (plateau in the right part of the figure).

These non-monotonicities do $not$ vanish if we take the lattices large
enough. They are part of the asymptotic scaling function and thus whatever
lattice size $L$ we take there will be a field $h \propto 1/L^{\beta \delta
/\nu}$ where $h/M^\delta$ has a maximum and near which it thus varies
non-monotonically with $L$.

We also tested the prediction of \cite{CD,CD2} that for fixed $T < T_c$ the 
leading finite-size deviation from the bulk value of $M$ should vanish 
exponentially in $L$, and not with a power law $\propto 1/L^d$ 
(as predicted by perturbation
theory based on the separation of the zero-mode [4-7]). Fig. 2a shows 
again non-monotonic behaviour, in both three and five
dimensions. But only in three dimensions these data are accurate enough
to distinguish between a tail varying exponentially and one $\propto 1/L^d$;
Fig.2b clearly supports the theoretically predicted \cite{CD,CD2} 
exponential variation. (These three-dimensional data were taken with Ito's 
fast algorithm \cite{ito}.)

In order to make contact with bulk properties we also used simulations at
the critical isotherm with $1292^3$ spins \cite{stauffer}. From the 
simulations we obtain the bulk value of $h/M^\delta$ as $D_c = 0.21\pm 0.02$. 
Our field-theoretic result in Eq. (28) is in very good agreement with this 
value. 

It is interesting to compare our simulation result also with other bulk 
theories. From series expansion Zinn and Fisher \cite{ZF} obtained 
$C^c = 0.299$ for the amplitude of the bulk susceptibility at the critical 
isotherm $\chi = C^c |h|^{-\gamma/\beta\delta}$  with $\gamma = 1.2395, 
\nu = 0.6320$. This leads to $D_c = 0.182$ according to the relation 
$D_c = (C^c \delta)^{-\delta}$. 

$D_c$ is also contained in the universal combination of amplitudes \cite{AH}
\begin{equation}
R_{\chi} = \Gamma D_c A_M^{\delta-1}
\end{equation} 
where $\Gamma$ is the amplitude of the bulk susceptibility 
$\chi = \Gamma t^{-\gamma}$ 
above $T_c$ at $h = 0$ and $A_M$ is the amplitude of the spontaneous 
magnetization $M_{bulk} = A_M |t|^\beta$ below $T_c$. Using $\varphi^4$ field 
theory  at $d = 3$ dimensions Guida and Zinn-Justin \cite{zinn} have obtained
$R_{\chi} = 1.649$. They also used the $\epsilon = 4-d$ expansion and obtained
$R_{\chi} = 1.674$ at $\epsilon = 1$. Using these values for $R_\chi$ and 
the high-temperature series expansion results \cite{LF} $\Gamma = 1.0928$, 
$A_M = 1.71$ and $\delta=(d\nu+\gamma)/(d\nu-\gamma)$ with $\gamma=1.2395, 
\nu=0.6335$, we obtain from Eq. (29) $D_c = 0.205$ ($d = 3$ field theory) and
$D_c = 0.202$ ($\epsilon$-expansion), respectively, in good agreement with
our theoretical result, Eq. (28), and with our MC simulation.

In summary, our simulations confirmed {\it a posteriori} our
theoretical predictions of Sect. 2 for the asymptotic finite-size effects 
in the three-dimensional Ising model. The agreement in Fig.1 is remarkable 
in view of the fact that the non-universal parameters of the theory were 
adjusted only to {\it bulk} parameters of the Ising model at $h=0$ and not 
to any finite-size MC data.

{\bf{Acknowledgments}}

Support by Sonderforschungsbereich 341 der Deutschen 
Forschungsgemeinschaft and by NASA under contract numbers 960838, 1201186, 
and 100G7E094 is acknowledged. X.S.C. thanks the National Science 
Foundation of China for support under Grant No. 19704005, D.S. thanks
the German Supercomputer Center in J\"ulich for time on their Cray-T3E.

\newpage

{\bf Figure Captions }

{\bf Fig. 1. } 
Scaling plot (in units of the lattice constant, see Ref.\cite{CDi}) of 
$h/M^{\delta}$ versus
$x=hL^{\beta\delta/\nu}$ below $T_c$ with $t/h^{1/\beta\delta}=-1.0$ in part a,
$t/h^{1/\beta\delta}= 0$ (that means $T=T_c$) in part b, 
$t/h^{1/\beta\delta}= 1.0$ in part c, and $t/h^{1/\beta\delta}= 1.6$ in part d.
Monte Carlo data for $L = 32$ and $80$.
Solid line is the theoretical prediction Eqs. (23)-(26); no Monte Carlo data 
are shown in part d, where $h/M^{\delta}$ is a monotonic function of external 
field $x=hL^{\beta\delta/\nu}$.

{\bf Fig.2. }
a) Monte Carlo data for the spontaneous magnetization in units of the lattice
constant in three (diamonds and plusses) and five 
(square) dimensions at $T/T_c = 0.99$, versus linear lattice size $L$. The
horizontal line $M = 0.3671$ for three dimensions is determined from $L = 2496$.

b) Selected three-dimensional data from part a) are shown as 
$M(L = \infty) - M(L)$ versus $L$ in a semilogarithmic plot. The straight 
line represents an exponential decay \cite{CD,CD2}, the two curved lines 
are power law decays $1/L^2$ and $1/L^3$ which fail to fit the data. 
The five-dimensional data of part a) were not accurate enough to distinguish 
between an exponential and a $1/L^5$ decay.

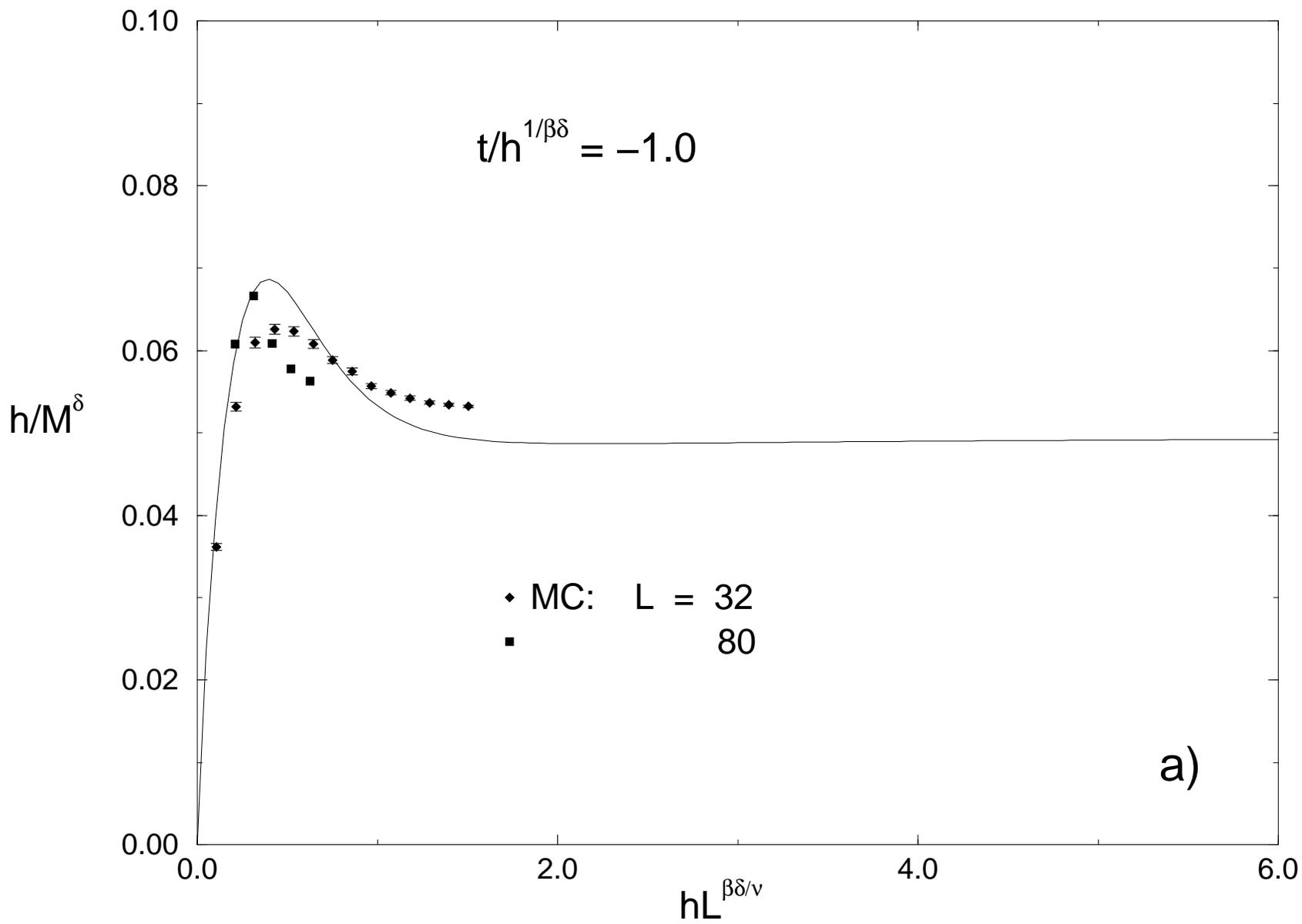
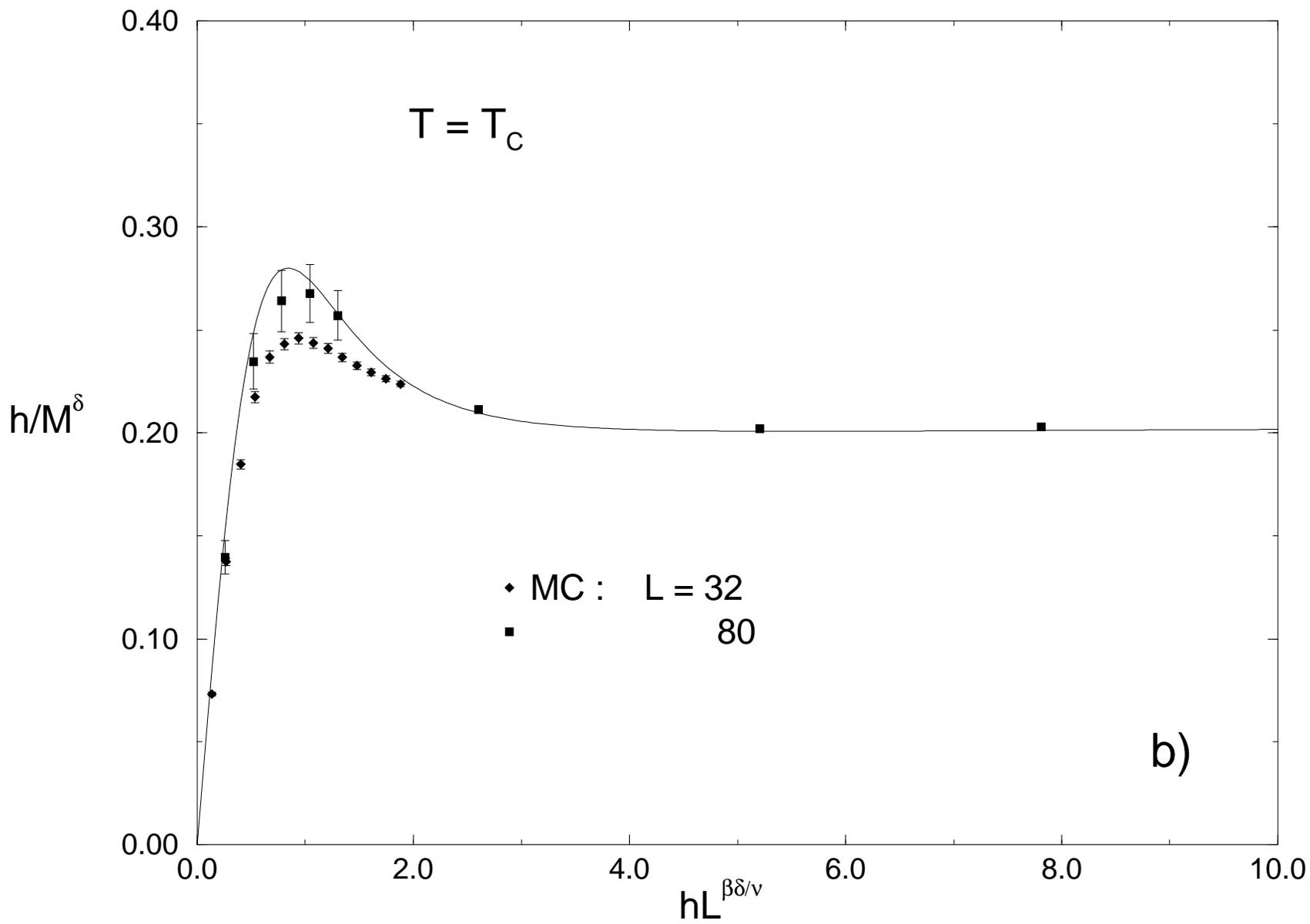
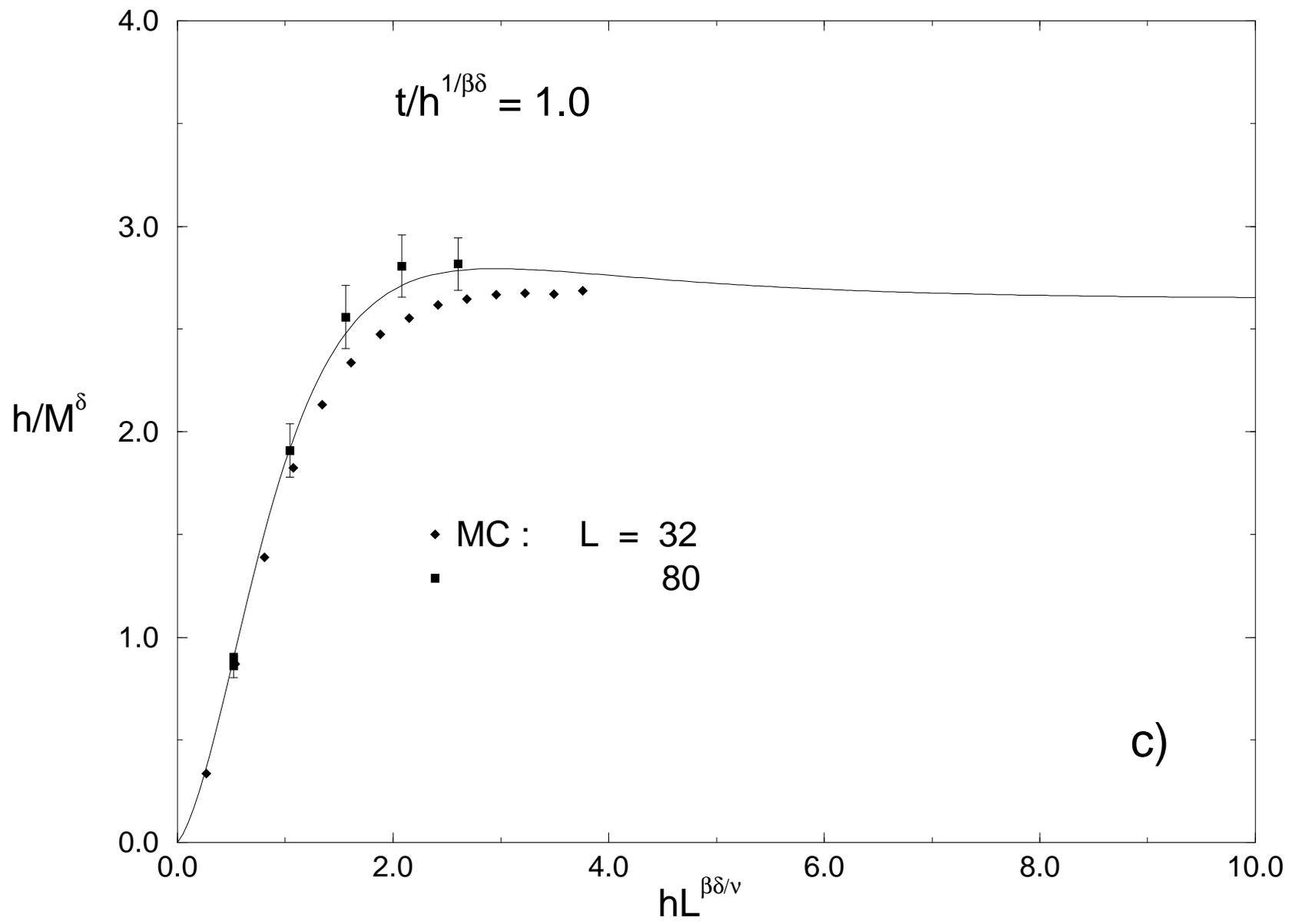
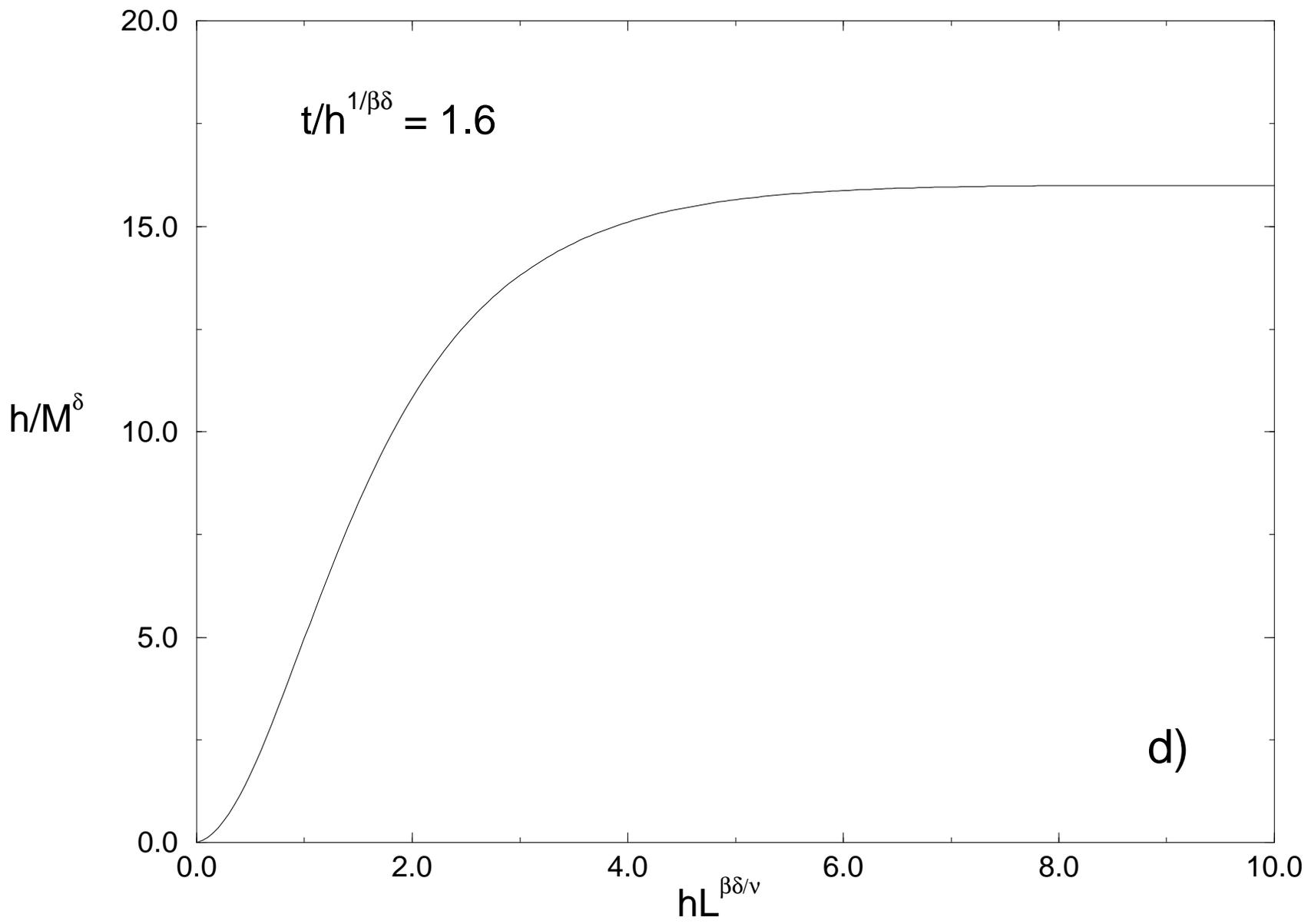
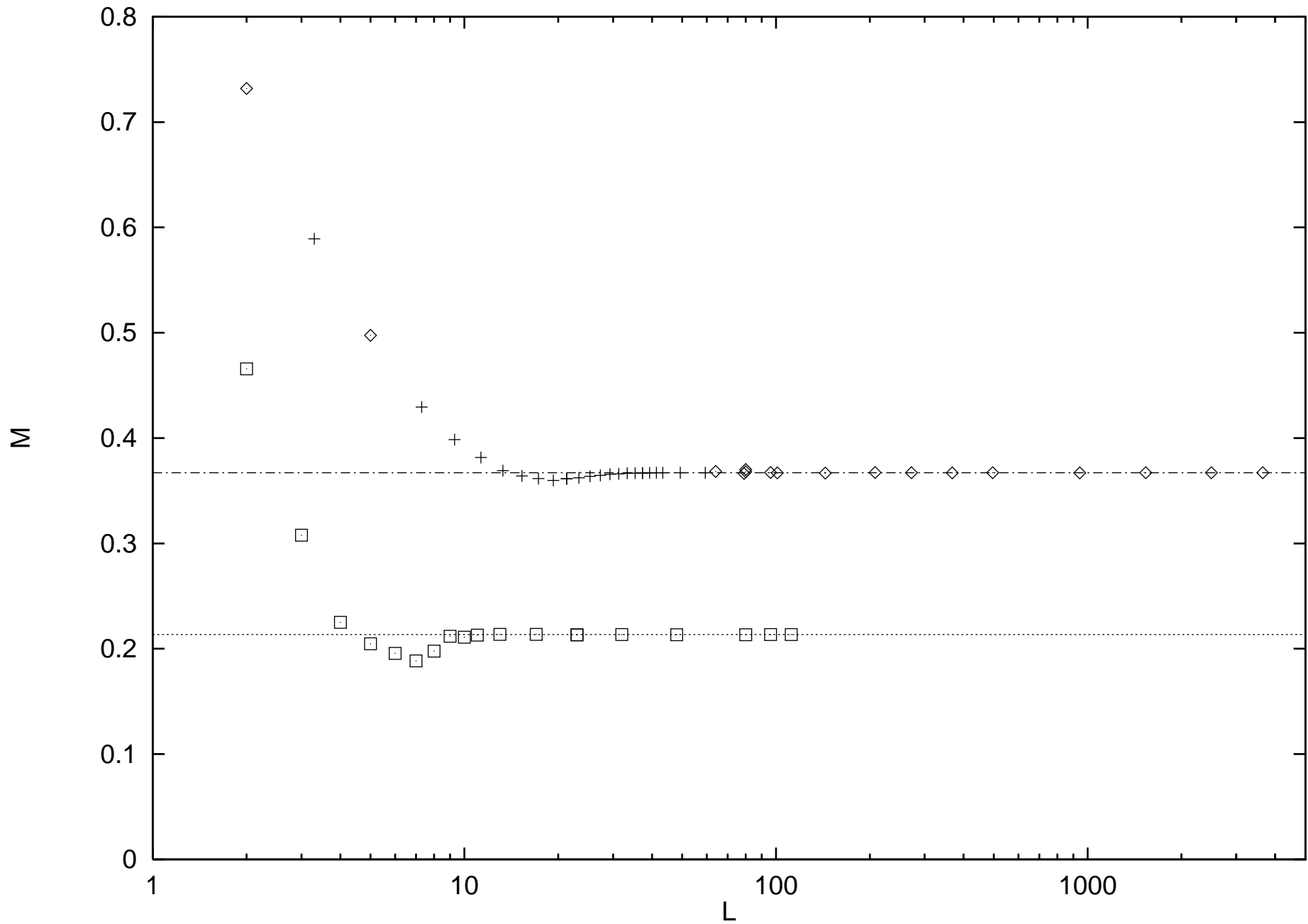
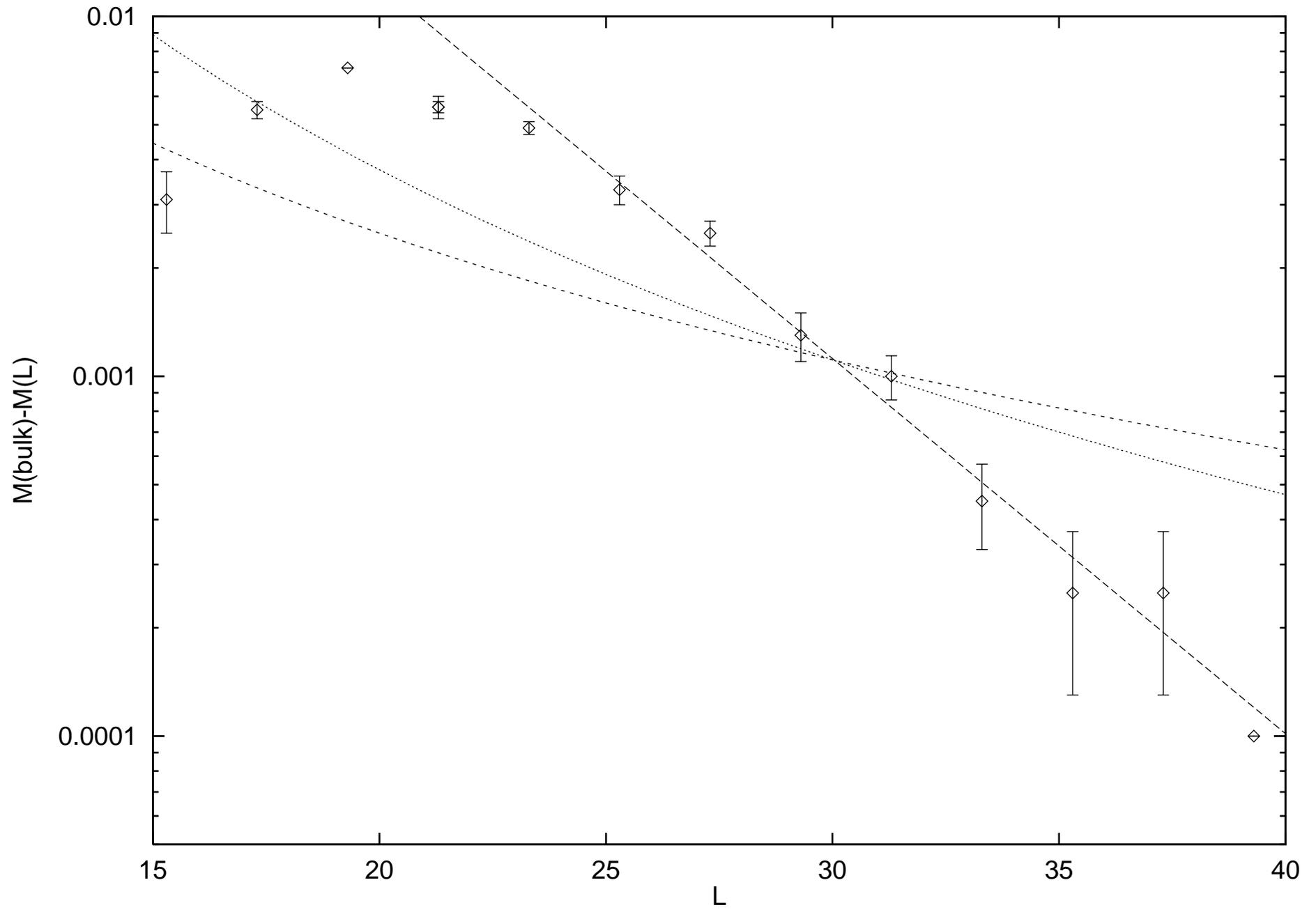
\end{document}